# MICROSTRUCTURED AIR-SILICA FIBRES: RECENT DEVELOPMENTS IN MODELLING, MANUFACTURING AND EXPERIMENT

Dominique PAGNOUX, Ambre PEYRILLOUX, Philippe ROY,
Sébastien FEVRIER, Laurent LABONTE, Stéphane HILAIRE

Institut de Communications Optiques et Microondes – Unité Mixte de Recherche n°6615
Faculté des Sciences et Techniques
123, Avenue Albert Thomas  F-87060  LIMOGES cedex

*Abstract:*
*The main modelling methods devoted to microstrutured air-silica optical fibres (MOFs) are presented and discussed. Then, the specific propagation properties of MOFs are studied in detail. Characteristics measured on fibres manufactured in our laboratory or reported in the literature are analysed. A large number of potential and demonstrated applications are presented and the obtained performances are discussed. A particular attention is given to hollow-core photonic bandgap fibres and their applications.*

LES FIBRES OPTIQUES MICROSTRUCTUREES AIR-SILICE: MODELISATION, FABRICATION ET EXPERIMENTATION

*Résumé:*
*Les principales méthodes de modélisation dédiées aux fibres microstructurées air-silice (FMAS) sont présentées et discutées. Puis les propriétés de propagation spécifiques des FMAS sont détaillées. Les caractéristiques mesurées sur les fibres fabriquées au laboratoire ou rapportées dans la littérature sont analysées. Un grand nombre d'applications potentielles ou démontrées sont présentées et les performances obtenues sont discutées. Une attention particulière est accordée aux fibres à cœur creux à bande interdite photonique et à leurs applications*

## I - INTRODUCTION

In the past years, guiding light into silica fibres using a transverse Bragg effect for maintaining light into the core has given rise to a large interest due to the potential unusual dispersion properties of such guides. First, so-called Bragg fibres with a cladding made of alternating high and low index layers have been proposed [1]. We have experimentally have demonstrated for the first time in 1998, that a single-mode propagation of light may be achieved into the core of a Bragg fibre, even if the index of the core is lower than that of the layers of the cladding [2]. This fibre was manufactured by the MCVD technique. Contrarily to conventional fibres, the extension of the electric field into the cladding of the Bragg fibre does not depend of the wavelength. The main consequence is that the dispersion of the guide may be kept positive at short wavelengths, and the zero dispersion wavelength may be shifted towards values lower than 1.28µm, and as short as 0.8µm [3]. The zero dispersion wavelength of the fibre described in [2] as been measured and is equal to 1.06µm, in good accordance with theoretical predictions. Because they are based on Bragg reflection, Bragg fibres also exhibit a strong band-pass filtering behaviour around the operating wavelength [4].

However, Bragg fibres manufactured with the MCVD technique suffer somewhat high propagation loss. Furthermore, this loss is dramatically sensitive to the curvature. The main reason for this is that the index contrast between the layers of the cladding remains small and the expansion of the mode into the cladding is considerable. The propagating field is then very sensitive to the imperfections of the index profile. In order to reduce this loss, a significantly higher index contrast between the layers of the cladding would be necessary, as it is achieved in a recently reported multimode Bragg fibre made with a thermal evaporation technique ($As_2Se_3$ n=2.8 / PES n=1.55) and operating in the 1-2µm range of wavelengths [5].

In 1995, P. St. J. Russel, T. Birks and their colleagues proposed an alternative technique for guiding light based on photonic band gap (PBG) structures [6,7]. Starting from the concept previously published by E. Yablonovitch [8], they theoretically designed a 2-D photonic crystal fibre (PCF) made of a periodic arrangement of micron-size cylindrical parallel air holes into a silica matrix. The most usual arrangements of the holes are with hexagonal symmetry, either triangular (centred-hexagonal) or of honeycomb type (non-centred hexagonal), as shown on figure 1a. The structure of the fibre is invariable by a translation along the axis of the holes (z-direction). To allow the transverse confinement of the light at a given place of the cross section of the photonic crystal, an opto-geometrical default is set at this place: either an additional air hole is set in the centre of an elementary mesh of the honeycomb arrangement, or one air hole of the triangular arrangement is replaced by silica (figure 1-b). The light then propagates along the z-direction into the region of the default that behaves like the core of the fibre. Let us notes that when the default is created by replacing an air hole by silica into a triangular lattice, the index of the core, which is that of pure silica, is higher than the mean index of the cladding because of the presence of air holes into the surrounding region. In this case, light may be simply guided thanks to the well-known total internal reflection (TIR) principle that operates into classical optical fibres.

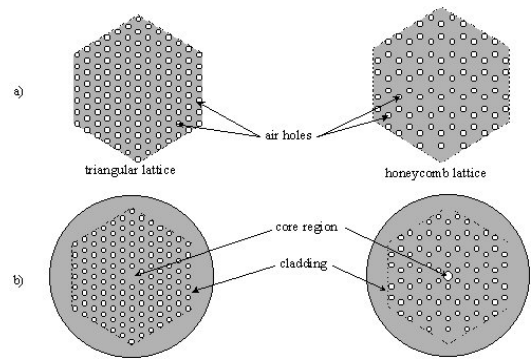

Figure 1: (a) cross section of a triangular type and honeycomb type photonic crystal structure; (b) cross section of a photonic crystal fibre based on a triangular type and a honeycomb type photonic crystal structure
*(a) section droite d'une structure à cristal photonique de type triangulaire ou à nid d'abeille; (b) section droite d'une fibre à cristal photonique basée sur une structure à cristal photonique de type triangulaire ou à nid d'abeille*

The very first attempt for manufacturing photonic crystal fibres based on a periodic lattice of air holes (the cladding) surrounding a defect of this lattice (the core) was reported in 1996 [9]. Even if the fibre was initially conceived for having a PBG effect into the cladding at 1.55µm, this effect was not observed. The light was simply guided thanks to the TIR principle.

In the literature, fibres made of an arrangement of air holes are wrongly often called "photonic crystal fibres" whatever the guiding principle they use. In the following, to avoid any confusion, we prefer not use this term. We shall distinguish the fibres versus their waveguiding principle, by employing the following specific terms for each: "microstructured air-silica optical fibres (MOFs)" will refer to fibres simply operating on the TIR principle, and fibres operating on the PBG principle will be called "PBG fibres".

Because of the non-conventional structure of the cladding, MOFs exhibit propagation properties very different to those of usual fibres in terms of spectral range of single-mode operation and chromatic dispersion, even if they simply operate on the TIR principle. These novel features were first observed by T.A. Birks et al. in 1997 [10]. Since this date, the potential performances of this new class of optical fibres have aroused a tremendous interest because they may have important applications in many domains such as telecommunications, non-linear optics and sensors. Intensive efforts have been paid in numerous laboratories all over the world for properly modelling, manufacturing and testing these fibres.

In this paper, we review recent progress reported on both MOFs and PBG fibres. In the first section, we focus on the different methods proposed for modelling MOFs and we briefly discuss their capabilities and limitations. Then, the attractive propagation characteristics measured on actual MOFs are reviewed. At last, we point out very promising results obtained with MOFs and PBG fibres and we describe novel applications, components and devices that follow.

## II - MODELLING METHODS

To deduce the waveguiding properties of MOFs, such as the range of single mode operation, the effective area of the fundamental mode, its chromatic dispersion and its propagation loss, an accurate electromagnetic study of the propagating fields must be achieved, providing their effective index versus wavelength and their transverse distributions. In this section, the most common methods are reviewed and discussed.

### II – 1  the effective V-model

The very first method for modelling the propagation of light into MOFs has been an effective V-model proposed by Birks et al. [10]. The aim was to define the opto-geometrical characteristics of a step-index fibre (SIF) having the same propagation properties as the considered MOF. The index of the core ($n_1$) is that of silica. The index of the cladding is the effective index of the fundamental space-filling mode transmitted in an infinite defectless lattice identical to that constituting the cladding of the MOF. This cladding index is noted $n_{SFM}$. Considering that this cladding is an assembly of hexagonal unit cells centred on one of the holes, the electric field of the equivalent SIF is described by Bessel functions with the proper boundary conditions. In this case $n_{SFM}$ is computed with:

$$n_{SFM}^2 = \frac{\iint n^2 |E|^2 dS}{\iint E^2 dS} - \frac{\iint \left|\frac{dE}{dr}\right|^2 dS}{k_0^2 \iint E^2 dS} \tag{1}$$

where E is the electric field, n is the refraction index of silica ($n_1$) or of the medium into the hole ($n_a$) versus the considered point, S is the area of an elementary cell, r is the distance to the centre of the fibre, and $k_0 = 2\pi/\lambda$ is the wave vector at the operating wavelength $\lambda$.

Because the extension of the electric field into the holes strongly depends on the wavelength, relation (1) shows that $n_{SFM}$ is an unusual function of the wavelength. To obtain a better description of the variations of $n_{SFM}$, one can introduce the air filling fraction f, which is given by:

$$f = \frac{\pi}{4}\left(\frac{d}{\Lambda}\right)^2 \text{ for a triangular lattice}$$

$$\text{or} \quad f = \frac{\pi}{6}\left(\frac{d}{\Lambda}\right)^2 \text{ for a honeycomb lattice} \tag{2}$$

where d is a hole diameter, and $\Lambda$ is the centre-to-centre spacing or pitch.
On the one hand, when the wavelength is large compared with d and $\Lambda$, the field largely spreads into the holes and the fundamental space filling mode can be considered as a plane wave propagating into an homogeneous medium having an effective index $n_{SFM}$ given by:

$$n_{SFM}^2 \approx n_1^2 - (n_1^2 - n_a^2)f \tag{3}$$

On the other hand, starting from the relation (1), one can show that for short wavelengths, an asymptotic expression of $n_{SFM}^2$ is [11]:

$$n_{SFM}^2 \approx n_1^2 - \frac{d}{\Lambda}\left(\frac{\lambda}{\Lambda}\right)^2 \tag{4}$$

The effective V-value of the MOF is then:

$$V_{eff} = \frac{2\pi}{\lambda} a_{eq} \sqrt{n_1^2 - n_{SFM}^2} \tag{5}$$

where $a_{eq}$ is the core radius of the equivalent SIF, that one must determine.

Considering that the core of the MOF consists in a defect in the air holes lattice, the precise localisation of the core-cladding boundary is somewhat hazardous and arbitrary. It was first proposed to set $a_{eq}=\Lambda$ [10]. In this case, to provide theoretical predictions in accordance with experiments, the cutoff frequency $V_c$ of the second mode is set at a value significantly higher than the usual value of 2.405 found for SIF. It is evaluated to be close to 4 [10,12]. However, with the help of further calculations including comparisons with results of other methods (see next section), we have shown that the propagation constants of the modes of the considered MOF are close to those of a SIF having a core radius $a_{eq}=0.64\Lambda$, for $d/\lambda>0.4$ [13]. When choosing this value of $a_{eq}$, the cutoff frequency of the second mode remains equal to 2.405.

Because of the strong variation of $n_{SFM}$ versus $\lambda$, the variation of $V_{eff}$ versus $1/\lambda$ is not linear, contrarily to that of classical SIF. Fig. 2 shows examples of curves of $V_{eff}$ versus the normalised parameter $a_{eq}/\lambda$ ($a_{eq}=0.64\Lambda$) for different ratios $d/\Lambda$. As it can be predicted with relation (4), in the limit of short wavelengths, the decreasing index difference cancels the increase of $1/\lambda$ in (5). Hence $V_{eff}$ has an upper bound lower than 2.405 if the ratio $d/\Lambda$ remains small enough (lower than about 0.4). This explains why MOFs with small enough air filling fraction are endlessly single-mode, at all wavelengths, even with a very large core. This feature has been experimentally demonstrated with a MOF having a 22µm core across dimension, at wavelengths as short as 450nm [14]. It is very attractive for high power laser transmission with reduced non-linear effects.

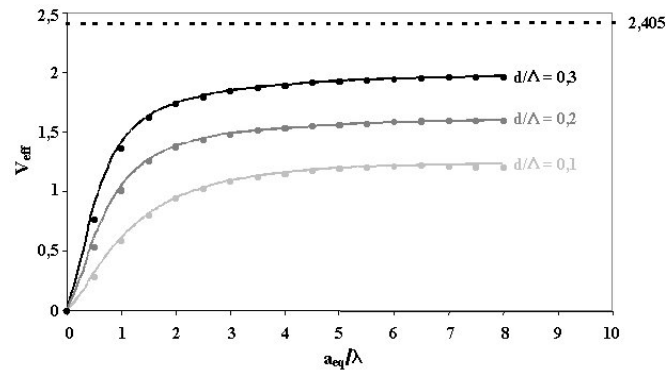

Figure 2: normalised frequency $V_{eff}$ plotted versus $a_{eq}/\lambda$, for different ratios $d/\Lambda$ : demonstration of the endlessly single-mode property of MOF with small enough air filling fraction
*Fréquence spatiale normalisée $V_{eff}$ tracée en fonction de $a_{eq}/\lambda$, pour différents rapports $d/\Lambda$ : démonstration de la propagation monomode illimitée dans une FMAS contenant une proportion d'air suffisamment faible*

The effective V-model only provides useful indications on the single mode range of operation of MOFs. For predicting their modal properties such as their dispersion or birefringence, it is obvious that other methods must be implemented.

**II – 2 the equivalent average index method (EAIM)**

We have proposed a very simple method, so-called "equivalent average index method", in order to provide an approximate but sufficiently reliable value of the effective index of the propagating modes into perfectly symmetric MOFs [15]. It is based on a development of the index profile and of the electric field into Fourier series, taking advantage of the π/3 symmetry of the fibre, and only considering the mean terms of these Fourier series. These mean terms, respectively $\overline{n^2(r)}$ and $\overline{E(r)}$, are given by:

$$\overline{n^2(r)} = \frac{3}{\pi}\int_0^{\pi/3} n^2(r,\varphi)d\varphi \qquad (6)$$

$$\overline{E(r)} = \frac{3}{\pi}\int_0^{\pi/3} E(r,\varphi)d\varphi \qquad (7)$$

where r and φ are respectively the radial and the azimuthal co-ordinates.

On Fig. 3, we have reported the two dimensional refractive index profile of a MOF (fig. 3a), and the averaged refractive index profile computed with (6). A classical resolution of the one-dimensional scalar wave equation provides the averaged electric field amplitude $\overline{E(r)}$ (fig. 3b) together with the associated propagation constant $\beta=k_0 n_e$. The chromatic dispersion $D_c$ is then computed with the well-known relation:

$$D_c = -\frac{\lambda}{c}\frac{d^2 n_e}{d\lambda^2} \qquad (8)$$

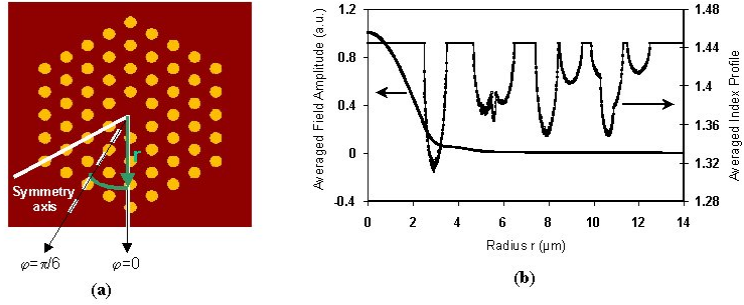

Figure 3: (a) two-dimensional refractive index profile of a MOF, (b) associated averaged refractive index profile $\overline{n(r)}$ and averaged electric field amplitude $\overline{E(r)}$

*(a) profil d'indice à deux dimensions d'une FMAS, (b) profil d'indice moyenné associé $\overline{n(r)}$ et amplitude de la valeur moyenne du champ $\overline{E(r)}$*

Because of its circular symmetry, the averaged electric field $\overline{E(r)}$ cannot fit the actual electric field of the MOF. However, the associated effective index $n_e$ and its second derivative, proportional to the chromatic dispersion, are in good accordance with those of the actual mode of the MOF, as long as the optical wavelength is not too enlarged [16]. This simplified method in then of great interest in order to roughly define the suitable opto-geometrical parameters of a MOF intended to a particular application, before using a more precise method for improving the results. Let us note that a close method has also been implemented for modelling modes of ring structures in microstructured polymer optical fibres [17].

## II – 3  the bi-orthogonal modal method (BOM)

This full-vectorial method was first proposed in 1998, in order to solve the modes of an inhomogeneous fibre. It is based on the non-self-adjoint character of the electromagnetic propagation in a fibre [18]. It leads to the algebraic determination of the solutions of a set of eigenvalue equations that involve the evolution operator relating the transverse components of the magnetic and the electric fields to the propagation constant of the modes.

For application to MOFs, one must first define a large superlattice including the core region and a large number of periodically set air holes, and exhibiting a periodicity of the structure following two orthogonal axes [19]. The main argument of the method is to impose that the sought fields fulfill periodic boundary conditions in the directions of these axes. These conditions allow the decomposition of the modal distribution of the fields into a plane waves basis, which transverse wave vectors are defined by the points of the reciprocal lattice associated to the superlattice. Thanks to this decomposition, the matrix elements of the evolution operator corresponding to a single hole can be obtained analytically. Moreover, taking advantage of the symmetry properties of the structure, the matrix of the evolution operator for the whole fibre can also be analytically carried out. A variational method, based on the same principle, has also been proposed for modelling true photonic bandgap fibres [20].

With this method, the analytical determination of the effective index of the modes allows to overcome the problem of loss of precision that can occur in numerical computations. The dispersion can be accurately computed and optimised (for example flattened) using the relation (8), by finely adjusting the opto-geometrical parameters of the fibre [21]. Such calculations show that the dispersion is very sensitive to these parameters: for example, a change of 0.01µm in Λ may induce a shift of ±1ps/(nm.km) of the dispersion around zero (figure 4). However, the method is only suitable for modelling MOFs having a perfectly symmetrical structure.

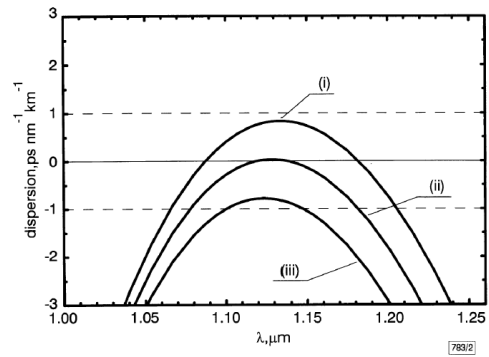

Figure 4: dispersion computed versus λ, for a ratio d/Λ=0.365 and Λ=1.74µm (i), 1.75µm (ii), 1.76µm (iii), starting from effective indices obtained with the biorthonormal-basis method (*A. Ferrando et al. [21]*.)
*Dispersion calculée en fonction de λ, pour un rapport d/Λ=0.365 and Λ=1.74µm (i), 1.75µm (ii), 1.76µm (iii), à partir des indices effectifs obtenus avec la méthode de la base bi-orthonormale (A. Ferrando et al. [21].)*

## II – 4  the localised functions method (LFM)

The LFM is based on the decomposition of the electromagnetic field and of the permittivity profile $n^2(x,y)$ into a sum of cosine and Hermite-Gaussian functions. It has first been proposed for applications to MOFs by Monro et al. in a scalar form [22] and then extended to a vector model [23]. These expressions of the electromagnetic field and of the index profile are introduced in the two-dimensional full vector wave equation. The resolution of this equation becomes an eigenvalue problem whose eigenvalues are the effective indices of each mode that might propagate in the fibre. The accuracy of this model depends on the number of Hermite-Gaussian functions. Typically, computations rapidly converge to a physical solution with about 100 functions for the field decomposition and for the permittivity decomposition.

This method has the advantage to be fast as it uses more analytical calculus than other methods, but it requires a lot of computational memory. The vectorial form of the model provides the mode field distribution and the effective index, even if the holes are not periodically arranged, as shown in figure 5 [24]. The effective area and the chromatic dispersion can be derived from these results. Even if the model is potentially able to take the polarisation into account, no theoretical study dealing with the birefringence of actual MOFs have been published yet with this method.

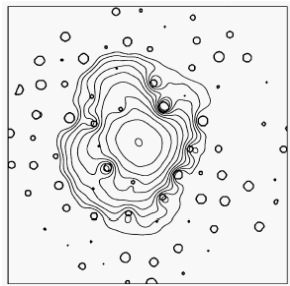

Figure 5: example of the mode field distribution into a MOF with random hole position and sizes (80 holes with d varying from 0 to 0.6µm, Λ≈2µm) *(T. Monro et al. [24])*
*Exemple de distribution du champ dans une FMAS ayant des trous de taille et de position aléatoires (80 trous avec d variant de 0 à 0.6µm, Λ≈2µm) (T. Monro et al. [24])*

## II – 5 the multipole method (MM)

A fully vector rigorous multipole method has recently been developed to explore MOFs properties. It is based on the use of scattering matrices of the holes into the fibre and on the translation properties of the Fourier-Bessel functions chosen to describe the electromagnetic field [25,26]. The method consists in finding homogeneous solutions of the scattering problem, i.e. in solving:

$$\mathcal{M} \, \tilde{B} = 0 \qquad (9)$$

where $\mathcal{M}$ is the appropriate scattering matrix and $\tilde{B}$ is a vector which associated field is a mode of the structure. Finding the modes for a given structure and wavelength thus leads to the search of the complex function $\det(\mathcal{M})$ of the complex variable $n_e$. In addition, the notion of generalised scattering matrix is employed for the external boundary of the fibre, so that the influence of the extent of the confining air-hole region on the dispersion or on associated loss of modes can be taken under consideration. The geometrical loss $L$ in dB/m associated with the modes is directly obtained from $J(n_e)$, the imaginary part of $n_e$ [27]:

$$L = \frac{20}{\ln(10)} \frac{2\pi}{\lambda} J(n_e) \times 10^6 \qquad (\lambda \text{ in } \mu m) \qquad (10)$$

The figure 6 shows an example of the dispersion and of the loss computed for a 3-rings MOF, versus wavelength for different hole diameters. One can note that flattened dispersion curves obtained for certain fibre configurations (small hole diameters) appear to be unsuitable for applications due to high loss that cannot be improved by a simple increase of the number of air holes rings.

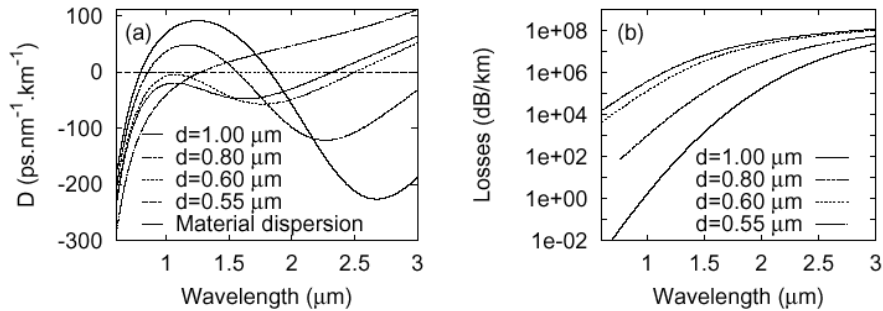

Figure 6: (a) dispersion and (b) loss of a 3-rings MOF as a function of wavelength and hole diameters d. The pitch $\Lambda$ is equal to 1.55μm (*by courtesy of G. Renversez*)
*(a) dispersion et (b) perte d'une FMAS à 3 anneaux de trous en fonction de la longueur d'onde et du diamètre des trous d. L'espacement $\Lambda$ est égal à 1.55μm (avec la permission de G. Renversez)*

As for the other mode-finders methods, the fundamental hypothesis of the multipole method is the invariance of the fibre along its axis. This method preferentially applies for MOF with circular holes, because the boundary conditions may be implemented analytically. However, for non-circular holes, differential or integral methods can be used to compute the scattering matrices associated with these holes [28]. Let us note that the method would no longer be appropriate if the circular regions including each hole would overlap, as it is the case into novel photonic crystal fibres with a air-core (refer to section IV).

## II – 6 the finite element method (FEM)

In order to model the propagating modes of MOFs, we have adapted a full-vector FEM previously developed in our laboratory for modelling and optimising microwave devices [29]. The cross section of the MOF is first split into a grid made of small elementary triangular subspaces. Then, the Maxwell equations are solved at each node of the grid [30]. The dimensions of each subspace of the grid are adapted to take into account the abrupt variations of the electromagnetic fields, especially at the boundaries of the holes close to the core. Typically, in order to preserve a suitable accuracy of the results, the largest dimension of the subspaces describing the core region is taken shorter than λ/5.

The FEM gives access to the distribution of the electromagnetic field vectors and to the associated effective permittivity for a fixed polarisation and a fixed frequency. By choosing the appropriate Magnetic Short Circuit and Electric Short Circuit (MSC-ESC) combination applied on the structure boundaries, the electromagnetic fields of the fundamental mode can be calculated using the FEM for two orthogonal polarisations. The field distribution for the two polarisations is different due to the effects of the conditions of continuity of the components of the field at the hole boundaries as shown in figure 7. At this point, one can notice that many models used to study MOFs find a perceptible difference between effective indices of orthogonal polarisations, as underlined in [31]. On the contrary, the same value of $n_e$ is found with the FEM, provided that the discretization of the fibre section is operated in such a way that the grid exhibits the same symmetry than the structure itself [32]. This result is in accordance with the demonstration of ref. [31] that states that a waveguide having a $2\pi/m$ rotation symmetry with m>2, such as a perfectly symmetric MOF, is not birefringent.

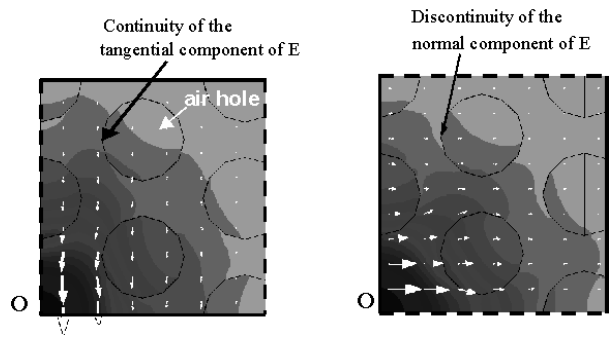

Figure 7: distributions of the electric field of the fundamental mode into the upper right quarter of the cross section of a MOF, for two orthogonal polarisations : these distributions are different because of the effects of the conditions of continuity of the components of the field at the hole boundaries ; however the effective index are found to be the same in the two cases
*Distributions du champ électrique du mode fondamental dans le quart supérieur droit d'une section droite d'une FMAS, pour deux polarisations orthogonales : ces distributions sont différentes du fait de l'application des conditions de continuité des composantes des champs aux interfaces avec les trous ; cependant les indices effectifs sont les mêmes dans les deux cas*

Thanks to the precise description of the field, the effective area can be accurately computed. The chromatic dispersion is also easily deduced from the $n_e$ values versus the wavelength. The reliability of the method has been demonstrated by numerous satisfactory comparisons of computed results with published experimental ones [30, 32-33].

When modelling waveguides with the FEM, the considered domain must be bounded. Generally, the outer limit of this domain is supposed to be a metallic surface operating as an electric short circuit, and making the electric field to be zero at the boundaries. The confinement loss of the structure can be evaluated by setting a lossy metal at the boundaries. This results in determining a complex propagation constant $\alpha + j\beta$ which real part provides the confinement loss.

One of the main drawbacks of the method is that a proper description of the field necessitates a lot of memory because the elementary subspaces of the discretization must be small regardless to the wavelength. When the considered cross section is large, the memory capacity of the computer can rapidly become insufficient. It is the reason why, in most cases, we must take advantage of the symmetries of the fibre index in order to limit the calculations to only one quarter of the fibre. Then, the method is not the most suitable for modelling MOFs with no symmetries such as those considered in [24].

### II – 7 the beam propagation method (BPM)

The BPM classically performs series of Fourier transform of the propagating field, followed by a multiplication by a proper propagation term in the direct and in the conjugated spaces, respectively [34]. This method necessitates that the index profile is previously discretized into elementary cells over which the index value is constant.

Few years ago, we have implemented this well-known method for the first time for studying the propagation of light into MOFs. It was implemented in its scalar weak-guidance form [35]. For example, we have observed the setting up of the fundamental mode into a single mode MOF. The power distribution into this mode after 0.5cm of propagation (excitation by a gaussian field) is in very good agreement with that obtained with the FEM (figure 8). However, because of the high-index contrast at the holes boundaries, a full vectorial BPM must be applied to properly deal with the vectorial nature of the field [36]. Even if the BPM cannot directly determine the modes of the structure, the effective index of each mode can approximately be evaluated with a method described in [34]. It is then possible to estimate the chromatic dispersion if the BPM is operated at different wavelengths [36].

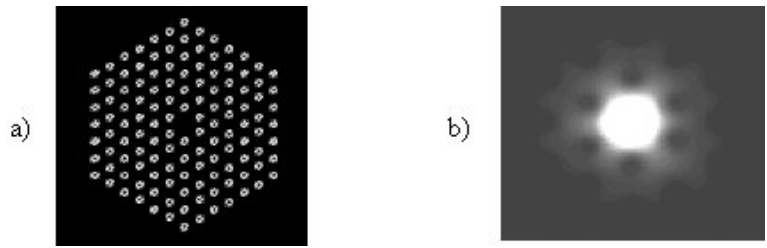

Figure 8: (a) discretization of the index profile of a MOF for modelling it with the BPM ; (b) power distribution into the fundamental mode (scale X3) after 0.5cm of propagation into the MOF
*(a) échantillonnage du profil d'indice d'une FMAS en vue de sa modélisation par la MFP ; (b) distribution de puissance dans le mode fondamental (échelle X3) après 0,5cm de propagation dans la FMAS*

Let us note that the BPM applied to optical fibres is time and memory consuming, because the spatial sampling step must be significantly lower than the wavelength for obtaining reliable results. Typically $\Delta x=\Delta y=\Delta z= 0.1\mu m$. Its main interest is that, contrarily to the previous mode-finders methods, it allows the study of the propagation into guides that suffer from longitudinal fluctuations of the index profile. In particular, the case of tapers, bends, couplers and splices can be considered.

**II – 8 discussion**

In this section, the most commonly used modelling methods for MOFs have been reviewed. Their main advantages and drawbacks have been pointed out. Among the mode-finders methods, full-vectorial ones must be preferred, in order to properly take into account the abrupt change of index at the hole boundaries. However, several ones are suitable only for MOFs exhibiting geometrical symmetries. Let us note that the choice of a method also depends on the searched characteristic. On the one hand, a precise value of the effective index is necessary for determining the single mode range of operation and for computing the chromatic dispersion. For evaluating the birefringence of a MOF, one must implement a method able to calculate the polarisation dependence of the effective index. On the other hand, the accurate determination of the mode field distribution is essential in order to calculate the effective area of this mode, which is useful for evaluating the power threshold of non-linear effects. Loss at splices with other fibres can also be evaluated knowing the mode field distribution. Some methods such as the multipole method or the FEM can compute the confinement loss of modes, versus the number of rings of holes into the cladding. At last, to study fibres suffering from longitudinal variations of their index profile, a BPM is clearly the natural choice, in spite of the very long computation time necessary to model a significant length of fibre.

Independently of these considerations, it is somewhat hazardous to compare the reliability of the different methods, because it largely depends on the opto-geometrical parameters of the MOFs and on the considered range of wavelengths. The speed and the available memory of the calculator also impose some limits to the implemented models, such as the size of the elementary subspaces for the BPM and the FEM, or the number of functions used with the LFM or the BOM. In the literature, one finds very few papers making comparisons of results obtained with different methods. In reference [36], the authors successfully compare the effective index deduced from their BPM to that obtained in [19] (d=1.2µm, Λ=2.3µm, λ varies from 300 to 1600nm). In [37], it is shown that the dispersion and the slope of the dispersion of a MOF (d=0.62µm, Λ=23µm, λ=0.81µm) provided by the multipole method are in good agreement both with our calculations published in [13] and with experimental results of [38]. In reference [39], we compare results provided by the FEM and the LFM. The range of agreement between the two methods is the same for the chromatic dispersion and its slope. We show that the discrepancies appear and increase as the pitch decreases. For a given pitch, the discrepancies are noticeable at small and high d/Λ values. They are attributed to the own imprecision of each method that induce more perceptible inaccuracy on the computed effective indices when the field is strong at the silica-air transitions (short pitches or high d/Λ values) or when the spread of the field over the cross section is important (short d/Λ values). An even more detailed analysis is achieved in [16], where we compare the mode field distribution and the dispersion of MOFs provided by the EAIM, the LFM and the FEM. These results are also compared with some published experimental measurements. We show that the FEM is reliable over a wide range of opto-geometrical characteristics. For its part, the EAIM does not manage very well with MOFs with small core diameter and small d/Λ ratio, especially at high optical wavelengths. That is to say that, as it could be easily predicted, the EAIM is not suitable for configurations leading to a strong azimuthal dependence of the electric field.

## III – FABRICATION AND MEASUREMENTS

### III – 1 fabrication of the MOFs

The principle used for manufacturing MOFs or PCFs is very similar to that of conventional fibres. The fibre is drawn at high temperature (around 2000°C) from a previously designed preform installed into a vertical furnace at the top of a drawing tower. The preform of MOFs consists in a careful arrangement of thin capillary silica tubes and rods stacked into a maintaining silica tube. The outer diameter of the capillary tubes is of about 1or 2mm, and the outer diameter of the preform measures a few centimetres. Most of the time, the drawn fibre has a typical outer diameter of 125µm for allowing easier connections with standard fibres. The general structure of the lattice of holes of the preform, i.e. the number and the disposition of these holes, is maintained into the fibre. However, one of the main differences with conventional fibres is that the optogeometrical parameters of a MOF do not systematically correspond to a homothetic reduction of those of the preform. Many parameters such as the temperature into the furnace, the pressure or the speed of the drawing have a significant influence on the size and on the shape of both the holes from the tubes and the interstitial holes. This is illustrated on figure 9a, where we show different cross sections of MOFs resulting from our very first manufacturing attempts, in 1998. One can first note that these fibres, drawn from the same preform with different manufacturing parameters, are clearly different. Furthermore, the unsuited manufacturing parameters induce large variations between the dimensions of the holes into one fibre. Some holes have even disappeared, when many interstitial holes have not collapsed. On figure 9b, we present some examples of cross sections of MOFs that we currently draw now in our laboratory, with the proper controlled manufacturing parameters. The structures are very regular and there are no more interstitial holes. Let us underline that, because of the strong dependence of the propagation characteristics of the MOFs on the diameter of the holes d and on the pitch Λ, it is absolutely necessary to perfectly control the manufacturing process in order to precisely obtain the desired performances.

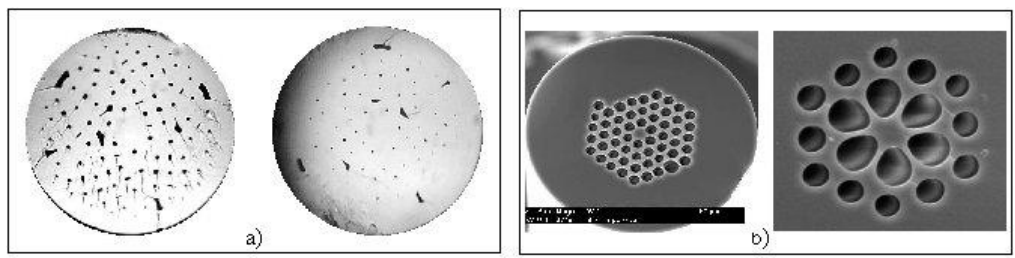

Figure 9: a) cross sections of different MOFs drawn from the same preform, with different unsuited manufacturing parameters ; b) cross section of MOFs drawn with the proper manufacturing parameters
a) sections droites de différentes FMAS étirées à partir de la même préforme, avec différents paramètres de fabrication inadaptés ; b) sections droites de FMAS étirées avec les paramètres de fabrication adéquats

The structures of the fibres essentially depend on the targeted applications. For example, strong non-linear effects may be obtained into a fibre with only three large holes (few tens of micron) surrounding a very small core (less than 2 microns) in which light is strongly confined (see section V). On the contrary, fibres with tens or hundreds of regularly arranged submicronic holes are suitable for telecommunications (see section III.2, III.5 and V). It is worth noting that, during the drawing process, the ratio between the preform and the fibre outer diameters is limited to about 200. Consequently, the fabrication of fibres with small pitches (smaller than 5μm) often necessitates two successive drawing steps during the process. A stack of 1-2mm in diameter rod and tubes constituting the preform is first drawn into a few millimetres in diameter microstructured capillary. This capillary is then inserted into a thick maintaining silica tube and the whole is drawn into the final fibre with the expected small pitch and hole diameter.

### III – 2  single mode propagation

One of the very first pointed out novel properties of MOFs was their ability to operate in a single-mode regime over a very large range of wavelength [10]. In this famous paper, Birks et al. experimentally proved the single-mode behaviour of a MOF (d=0.5μm, Λ=2.3μm) between 0.34μm and 1.55μm. The physical explanation of this unusually large single-mode range is developed in the section II – 1 devoted to the effective V-model. However, when the air-filling fraction of MOFs is increased for ensuring a robust guiding of light, the numerical aperture (NA) is consequently enhanced and the cutoff wavelength for the second mode is logically increased. For example, the fibre fabricated at Alcatel, which cross section is shown on figure 10 (d= 1.8μm, Λ= 2.4μm), is single mode at λ=1.55μm, but it guides the second mode at λ=0.98μm.

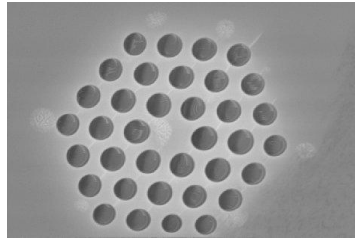

Figure 10: Cross section of a MOF (d=1.8μm, Λ=2.4μm) single mode at 1.55μm and bimode at 0.98μm
*Section droite d'une FMAS (d=1.8μm, Λ=2.4μm) monomode à 1.55μm et bimode à 0.98μm*

### III – 3  propagation loss

Propagation loss of light into MOFs is due to the sum of the confinement loss owing to a finite number of air holes that can cause the mode to be leaky, material loss and scattering loss. The confinement loss may be theoretically evaluated, assuming that the other causes of loss are zero, by means of some methods among which the BPM, the FEM and the multipole method. This confinement loss increases as the extension of the field increases, i.e. as the wavelength increases and/or as the ratio d/Λ decreases. For example, at λ=1550nm, a MOF with 3 rings of holes is subject to a confinement loss of 2dB/km (d= 1.15μm, Λ=2.3μm) and $10^4$ dB/km (d= 0.46μm, Λ=2.3μm) [40]. Then, one must ensure that the number of rings of holes is sufficient to make this kind of loss negligible.

Up to a recent date, MOFs have suffered from somewhat high optical loss, from 50 to 200 dB/km at λ=1550nm [41]. Assuming no confinement loss, this loss is attributed to many causes: impurities into the silica used for making the preform, surface roughness of the silica tubes and rods, contamination by OH ions during the drawing process, changes of holes diameters and/or hole pitches along the fibre due to an insufficient control of the manufacturing parameters. In 2002, Tajima et al. have demonstrated that, by minimising these causes, the loss can be decreased to 2dB/km at λ=1300nm and 1dB/km at λ=1550nm into the MOF of figure 11a comprising a silica-core surrounded by 4 rings of holes  (d=1.6μm, Λ=2.4μm) [42]. In the same time, Farr et al. obtained an even reduced loss (0.61dB/km at λ=1550nm) into the MOF, which cross section is represented on figure 11b (d=1.4μm, Λ=3.2μm). It was manufactured by implementing an improved fabrication process intended to reduce the effects of OH ions and other contaminants [43]. For reducing the losses associated with the evanescent fields within the holes, it is necessary to reduce the evanescent tail of the guided mode into these holes. To reach this goal, a particular type of MOF, so-called "hole-assisted lightguide fibre (HALF)" has been proposed: it consists in a core made of doped silica for enhancing the index and waveguiding, surrounded by pure silica with air holes for modifying the optical properties [44]. In the HALF which cross section is depicted on figure 11c (central $GeO_2$ doped core $\Delta n=6.10^{-3}$ and diameter of 7.4µm, 4 holes having a diameter of 15µm), the loss is of 0.73dB/km at $\lambda$=1550nm.

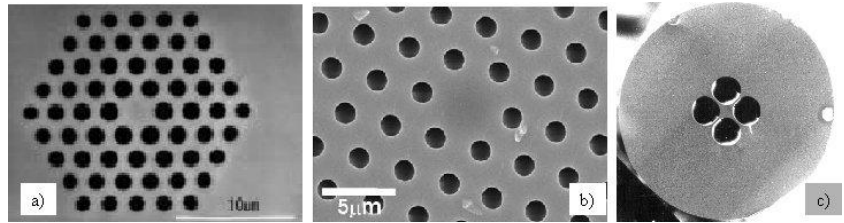

Fig 11 : Cross sections of low loss MOFs. At $\lambda$=1550nm, the loss is: a) 1dB/km (*Tajima et al. [42]*), b) 0.61dB/km (*Farr et al. [43]*), c) 0.73dB/km (*Hasegawa et al. [44]*)
*Sections droites de FMAS à faibles pertes. A $\lambda$=1550nm, les pertes sont: a) 1dB/km (Tajima et al. [42]), b) 0.61dB/km (Farr et al. [43]), c) 0.73dB/km (Hasegawa et al. [44])*

### III – 4  bend loss

Just like standard fibres, MOFs are very sensitive to bend loss at long wavelengths because the field more and more spreads into the cladding as the wavelength increases. A much more specific characteristic of MOFs is that they are also highly sensitive to bend loss at short wavelengths. This due to the fact that the core-cladding index difference decreases as the wavelength decreases. In order to define some operational limits, one can adapt the well-known formulae established for bend loss in conventional fibres with the aid of the effective index model [10]. One of these formulae indicates that, at short wavelengths, the critical radius of curvature $R_c$ under which the bend loss becomes unacceptable is:

$$R_c \propto \Lambda^3 / \lambda^2 \qquad (11)$$

MOFs with large core (diameter $2\Lambda$) have consequently a large pitch $\Lambda$ and are more sensitive to bend loss, as suggested by the relation (11). This behaviour has been clearly verified experimentally [10, 12]. Incidentally, the strong dependence of bend loss of certain MOFs versus wavelength has been observed into a MOF made of several silica cores *surrounding* a silica region with many small holes: when white light is coupled into the cores of a 1m sample, bending the fibre results in a colouring of the light emerging at the output end [45].

For the above reasons, MOFs exhibit both upper and lower bend edge (with respect to wavelength). Between these two limits, there is a range of wavelengths in which the bend loss is particularly low. For example, in our laboratory, we have measured the bend loss of the MOF shown in the inset to figure 12 that has been fabricated at Alcatel R&I in Marcoussis (d= 1.9µm, $\Lambda$= 2.3µm). The loss remains lower than 0.2dB, for ten turns of the fibre around a coil, whatever the diameter of the coil in the range 0.5cm – 4cm, and whatever the wavelength from 0.633µm to 1.5µm.

Considering the large number of optogeometrical parameters defining the structure of a MOF, it is difficult to draw a general conclusion of our and other published measurements. However, Broeng et al. state that MOFs with a pitch approximately twice the operational wavelength and a ratio d/$\Lambda$ of at least 0.25 are more bending resistant than standard optical fibres [46]. Then, bend loss does not constitute an obstacle to the use of MOFs for any application, provided that the other propagation characteristics fit with the target ones.

### III – 5 chromatic dispersion

One of the most attractive features of MOFS is the possibility that they offer to finely manage their chromatic dispersion. On the one hand, into MOFs having small air holes, the influence of these holes on the field is weak and the dispersion curve is close to the material dispersion of the silica. On the other hand, calculations show that the waveguide dispersion can be strongly increased as the size of the holes is increased. This implies that the waveguide dispersion of the fundamental mode may be positive at wavelengths lower than the zero dispersion wavelength of silica (1.28µm). It is then possible to cancel the chromatic dispersion of a single-mode MOF even at short wavelengths. This property, very attractive for continuum generation or soliton propagation, had been demonstrated before, both theoretically and experimentally, only into Bragg fibres [47,48].

Many works dealing with modelling the group velocity dispersion (GVD) of MOFs have already been published. Starting from the calculation of the effective index of the propagating mode versus wavelength, performed by means of one of the methods listed in section II, the GVD is derived from the relation (8). In the literature, many potential attractive features of the GVD into specifically designed MOFs have been theoretically demonstrated. The first ones concern the flattened GVD over a wide range of wavelengths: GVD comprised within $\pm 1$ps.nm$^{-1}$km$^{-1}$ for 1.05µm<$\lambda$<1.22µm (d= 0.636µm, $\Lambda$= 1.72 µm) [21] ; GVD comprised between 0 ps.nm$^{-1}$km$^{-1}$ and 10 ps.nm$^{-1}$km$^{-1}$ for 1.25µm<$\lambda$<1.9µm (d= 0.69µm, $\Lambda$=2.3 µm) [22] ; nearly zero ultra-flattened GVD over a very large range of wavelengths: the GVD may lie within (-0.5, +0.5) ps.nm$^{-1}$km$^{-1}$ for 1.31µm<$\lambda$<1.74µm into a MOF where d exactly equals 0.548µm and $\Lambda$=2.3µm [49]. In our laboratory, different MOFs have been designed for obtaining a flat GVD in the 1550nm range of wavelengths. For example, we have shown that the chromatic dispersion of a MOF with d=0.696µm and $\Lambda$=2.784µm can be maintained equal to 12$\pm$0.026 ps.nm$^{-1}$km$^{-1}$ in the C-band. For this fibre, $D_c$=12$\pm$0.55ps.nm$^{-1}$km$^{-1}$ over the 1480nm-1610nm range of wavelength.

Generally speaking, the methods for measuring the GVD are based on two basic principles: interferometric measurements on short samples of fibres and time flight measurements in long fibres spans [50]. Because of the difficulties of manufacturing long lengths of homogeneous MOFs and because of the somewhat high propagation loss in them, the first kind of technique is preferred for characterising MOFs. White-light interferometry [51] or low-coherence interferometry [52] may be implemented. To our knowledge, very few experimental results on flattened GVD have not been published yet. However, two MOFs showing dispersion of 0$\pm$0.6ps.nm$^{-1}$km$^{-1}$ from 1.24µm to 1.44µm, and 0$\pm$1.2ps.nm$^{-1}$km$^{-1}$ over 1µm-1.6µm respectively have been measured [53]. In order to achieve a sufficient confinement, these fibres are made of 11 rings of holes between the core and the external jacket (d= 0.58µm, $\Lambda$=2.59µm). The main applications of these features are obviously the high bit rate optical communications. However, the reliability of these fibres could be limited by the dramatically sensitivity of their dispersion curves to little changes in the optogeometrical parameters, as indicated in different papers and corroborated by our calculations.

Other interesting results deal with the cancellation of the GVD at short wavelengths. Concerning this feature, several experimental results have been published. In our laboratory, the GVD of the MOF shown in the inset to figure 12 (d=1.9µm, $\Lambda$=2.3µm) has been measured by a technique based on the low-coherence interferometry. The GVD is equal to about 130ps.nm$^{-1}$km$^{-1}$ at $\lambda$= 1.55µm and to 90ps.nm$^{-1}$km$^{-1}$ at $\lambda$=1.06µm (figure 12). It is cancelled at a wavelength lower than 800nm. The most significant experimental results published elsewhere are: for a MOF (d= 0.6µm, $\Lambda$=2.3 µm), GVD= -77 ps.nm$^{-1}$km$^{-1}$ at $\lambda$=813nm, instead of -110ps.nm$^{-1}$km$^{-1}$ for a standard fibre [54, 51] ; cancellation of the GVD at $\lambda$=740nm into a MOF (d=1.5µm, $\Lambda$=1.7µm) when the material dispersion of silica is about -130 ps.nm$^{-1}$km$^{-1}$ at this wavelength [55, 52] ; cancellation of the GVD at $\lambda$=560nm into a MOF with a very small core (diameter of 0.5µm) surrounded by a web comprising holes about 50 larger than the core [56] ; cancellation of the GVD at $\lambda$=1064nm into a MOF with d=1.1µm and $\Lambda$=2.5 µm (GVD = 46 ps.nm$^{-1}$km$^{-1}$ at 1550nm, and close to that of silica in the visible region) [57, 52]. This last result is clearly of great interest for non-linear operations requiring phase matching at the Nd:YAG laser wavelength. Concerning the hole-assisted fibres, a large anomalous dispersion of 43.8 ps.nm$^{-1}$km$^{-1}$ at $\lambda$=1550nm has been measured in the HALF shown on figure 11c.

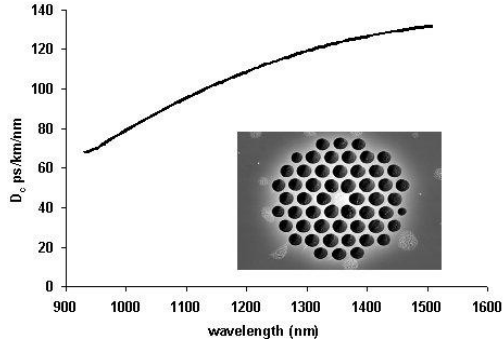

Figure 12: Experimental curve of the GVD of the fibre which cross section is shown in the inset (d=1.9μm, Λ=2.3 μm)
*Courbe expérimentale de la DVG d'une fibre dont la section droite est représentée dans l'encart (d=1.9μm, Λ=2.3 μm)*

### III – 6  birefringence

It is well-known that a waveguide having a rotational symmetry corresponding to a rotational angle $=2\pi/m<\pi$, is isotropic [31]. Thus, in the case of MOFs with a regular hexagonal arrangement of air holes (m=6), the birefringence, which is the difference between the effective indices of orthogonal polarisations, must be equal to zero. However, some birefringence can appear into such a structure, because of the residual internal stress of the material. Different theoretical works have been carried out for evaluating the birefringence due to perturbations in the holes diameters and/or in their positioning. For example, it has been shown that the birefringence of a MOF increases with the wavelength. For a MOF with d=1.03μm and Λ=2.3μm, it reaches values from $2\ 10^{-5}$ (at $\lambda=0.633\mu m$) to $1.5\ 10^{-4}$ ($\lambda=1.55\mu m$) when the holes diameters are randomly varied within ±1% and their distance to the axis is varied within ±0.5% [58]. The birefringence is likely to increase for a decreasing size of the structure. Another theoretical work achieved by means of the FEM results in the conclusion that the birefringence of a MOF (d=1.4μm, Λ=2.3μm) increases from $1.7\ 10^{-4}$ (at $\lambda=1.25\mu m$) to $3.5\ 10^{-4}$ ($\lambda=1.65\mu m$) when the diameters of all the holes of the two rings around the core, but those a symmetry axis, are enlarged by 10% [59]. These significant values indicate that a particular attention must be paid when manufacturing MOFs with an expected low birefringence.

In [32], we report the measurement of the birefringence of the MOF shown in the inset to figure 13 (d=0.5μm, Λ=2μm), achieved by means of an accurate magneto-optical method (see figure 13) [60]. Although no obvious geometrical dissymmetry is observed on the cross section of this fibre, we have measured an appreciable birefringence in the visible and in the near IR spectrum: a linear decrease versus $1/\lambda$ from $2.5\ 10^{-5}$ at $\lambda=1550nm$ (beat length = 6.8cm) to $2.5\ 10^{-6}$ at $\lambda=633nm$ (beat length = 25.7cm). The increase of the birefringence with the wavelength, also reported in [58], is contrary to the behaviour observed in standard fibres. We attribute this particularity to the fact that the accidental birefringence is expected to increase when increasing the difference between effective indices of the core and of the cladding, i.e. when the wavelength is increased.

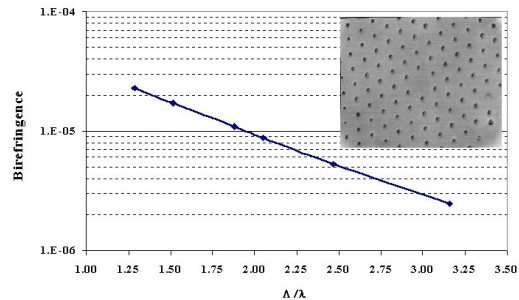

Figure 13: Birefringence of the fibre shown in the inset (d=0.5μm, Λ=2μm), measured by a magneto-optical method
*Biréfringence de la fibre représentée dans l'encart (d=0.5μm, Λ=2μm), mesurée par une méthode magnéto-optique*

Other measurements utilising a coherent reflectometry method [61] concern the MOFs with large cores shown on the figures 14a to 14c [62]. As predicted, the birefringence is somewhat lower than that we have measured, because of the larger size of the considered structures (see the optogeometrical parameters in the caption of fig.13). At λ=1550nm it is equal to 2.1 $10^{-6}$ (beat length=73cm), 7.4 $10^{-8}$ (beat length=21m), and 0.97 $10^{-6}$ (beat length=1.6m), for PCF10, PCF15 and PCF20 respectively. The very small birefringence of PCF15 is explained by the very good symmetry of this fibre. Even if low birefringence can be obtained into MOFs, their manufacturing process is much more favourable to the realisation of highly birefringent fibres. Indeed, it is easy to introduce a twofold rotational symmetry using air holes of different sizes disposed symmetrically on both side of the pure silica core. An example of such a fibre is shown on figure 14d, (small holes d=0.4μm, large holes d=1.16μm, Λ≈2μm) [63]. Its birefringence is as high as $4.10^{-3}$ (beat length=0.56mm at λ=1540nm). In our laboratory, we have designed a MOF of this type with a birefringence as high as 3.9 $10^{-3}$ which has been experimentally confirmed by a measurement of a beat length as short as 0.4mm at λ=1.55μm.

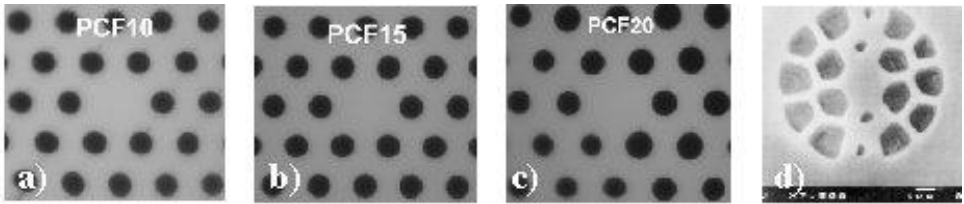

Figure 14: a, b, and c ) cross sections of MOFs with large core and reduced birefringence  (PCF10: d=2.5μm and Λ=5μm ; PCF15: d=3.4μm and Λ=6.8μm ; PCF20: d=4.5μm and Λ=8.9μm) (*Niemi et al. [62]*) ; d) cross section of a highly birefringent MOF (*Ortigosa-Blanch et al.[63]* )
*a, b, et c ) sections droites de FMAS à large cœur et biréfringence réduite  (PCF10: d=2.5μm and Λ=5μm ; PCF15: d=3.4μm and Λ=6.8μm ; PCF20: d=4.5μm and Λ=8.9μm) (Niemi et al. [62]) ; d) section droite d'une FMAS fortement biréfringente (Ortigosa-Blanch et al.[63] )*

### III – 5  unconventional effective mode area

At an operating wavelength, the spatial distribution of the mode may be adjusted into a MOF by properly choosing the dimensions of the holes and the pitch. Its knowledge allows the calculation of the effective mode area $A_{eff}$ that is an important parameter because it has connections with the numerical aperture, leakage loss, loss at splicing with other fibres, and above all non-linear effects [64]. $A_{eff}$ may be computed by means of the following relation:

$$A_{eff} = \frac{\left( \int\int_{-\infty}^{+\infty} |E(x,y)|^2 \, dxdy \right)^2}{\left( \int\int_{-\infty}^{+\infty} |E(x,y)|^4 \, dxdy \right)} \quad (12)$$

Our measurements of the field distribution into the MOF of figure 13 leads to values of $A_{eff}$ in good agreement with our calculations carried out by means of the FEM (figure 15).

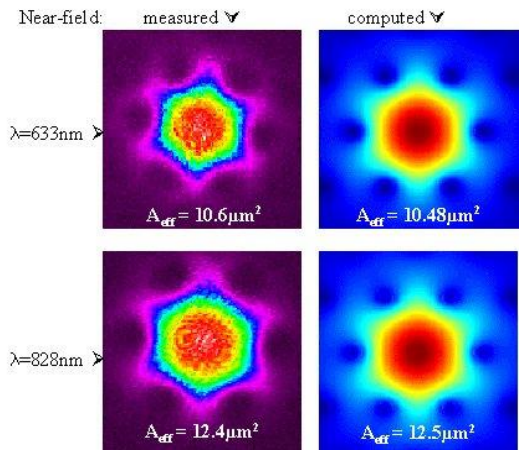

Figure 15: Near field pattern at the output of the MOF of figure 13 at two different wavelengths ; measured computed values of the effective area
*Champ proche à la sortie de la FMAS montrée figure 13 à deux différentes longueurs d'onde ; valeurs de l'aire effective mesurées et calculées*

As it as already been pointed out in this paper, a MOF may remain single-mode, even with a very large mode area. This feature is of great interest for propagating optical power with reduced non-linear Kerr effects, such as in long-haul optical communications links. One of the very first reported large mode area MOF was a fibre with d≈1µm, Λ=11µm and a core diameter of 22µm [65]. In spite of this very large core, the fibre is single-mode at a wavelength as short as 458nm. The MOFs shown on figures 14a to 14c, with a much larger air fraction, have cores as large as 19µm and remain single-mode at λ=1550nm [62].

On the contrary, MOFs with small cores and very large air-filling fractions offer the opportunity to confine light tightly within the core. This results in significant non-linear effects induced by modest optical powers, thanks to higher power densities. Thus, many applications dealing with non-linear optics, such as frequency generation or soliton propagation, can be considered. MOFs with effective area as small as few µm$^2$ have already been demonstrated. Ranka et al. propose a MOF made of a classical hexagonal arrangement of large holes (d=1.3µm) around a silica core as small as possible (diameter ≈1.7µm), as shown on figure 16a [66]. In our laboratory, we have drawn a preform in the centre of which three capillary tubes with large holes are stacked. The drawing conditions are set in order to collapse the interstitial hole between these tubes, this region becoming the core of the fibre (figure 16b) [67]. A similar technique is used in [68] to manufacture a MOF having a 2µm core, with $A_{eff}$ equal 3µm$^2$ at λ=1550nm (figure 16c).

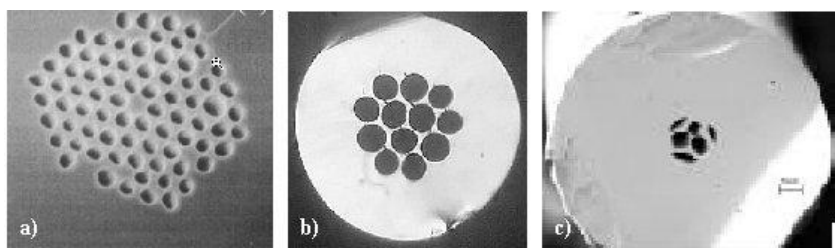

Figure 16: cross sections of MOF with small effective areas: a) MOF made by a classical hexagonal arrangement (*Ranka et al.[66]*) ; b) MOF manufactured in our laboratory ; c) MOF fabricated at ORC-Southampton (*Monro et al.[68]* )
*Sections droites de FMAS à petites aires effectives: a) FMAS fabriquée avec un arrangement hexagonal classique (Ranka et al.[66]) ; b) FMAS fabriquée dans notre laboratoire ; c) FMAS fabriquée à ORC-Southampton (Monro et al.[68] )*

# IV– RECENT PROGRESS IN PHOTONIC BANDGAP FIBRES

In two-dimensionally periodic materials, there can be ranges of the propagation constant normal to the periodic plane (i.e. the axial propagation constant $\beta$) where propagation is forbidden [69]. This means that light in certain wavelength bands cannot propagate through this photonic band gap (PBG) material. When light in this wavelength band is launched into a defect created in the centre of the PBG material, it is trapped into this defect and it can be guided in a single mode along the invariant axis of the structure. MOFs operating on the PBG principle have a cross section such as that depicted in figure 1a. The cladding is a 2D-PBG material, so-called photonic crystal, made of a near-perfect array of air holes with lattice constants of the order of micrometers, running down the length of the fibre. The core is a defect in the structure that may be an additional air hole. In this case, in this so-called "PBG fibre", the light can be strictly guided, without leakage, in a hollow core. Among the potential applications of single mode vacuum waveguides, the most exciting are ultra high power transmission, and the guiding of cold atoms.

The bandgaps and the propagating modes of PBG fibres can be theoretically determined by means of the full-vectorial supercell plane wave method [70, 20]. The very first experimental demonstration of air-guiding into a PBG fibre has been reported by Knight et al. in 1998 [71]. The fibre consists in a honeycomb pattern of air holes with an extra air hole in the centre. A blue-green single mode pattern was observed at the output end of a 5cm long fibre excited by a white light beam. This experiment first shows that very slight variations of the structure or scale of the fibre cause dramatically changes in the waveguiding properties, or can even make it disappear. A careful observation of the near-field pattern points out that in honeycomb PBG fibres, the guided mode is substantially distributed in silica, inducing a significant irreducible loss. This constitutes a fundamental limitation of the honeycomb design. A much more promising structure for concentrating the optical power in air is the use of a triangular based lattice of holes for the cladding around a large hollow core obtained by omitting the central seven tubes from the preform. Such a fibre was first manufactured by Cregan et al. in 1999 (d=3.5-7µm, Λ=5-10µm, air-filling fraction ~ 45%) [72]. Even if propagation of light was only demonstrated over a length of fibre as short as 3cm, the central air core was filled with a single lobe and the field intensity felt to very low values at the glass-air boundary. The loss is essentially due to an insufficient axial uniformity of the fibre structure. Since this first performance, extraordinary progress has been achieved in different laboratories. Transmission of light was obtained over a 15m long PBG fibre optimised for transmitting in the infrared (d=2.1µm, Λ=2.3µm, air-filling fraction ~ 80%) [73]. 3 transverse modes were identified into the fibre, each one propagated at one particular wavelength corresponding to its Bragg resonance into the cladding. However, the propagation loss remained still very high. Since this experiment, the loss has been considerably reduced. West et al. first reported, in 2001, an attenuation of 1dB/m at $\lambda$=1300nm [74]. Then, the year after, researchers from the same group have obtained a loss as low as 13dB/km at $\lambda$=1500nm, and less than 30dB/km over the transmission window between 1395nm and 1520nm, into the PBG fibre shown on figure 17a [75]. The loss measured around 1450nm, over a 100m sample of this fibre, is plotted on figure 17b. The spectacular decrease of the loss is mainly attributed to an excellent uniformity of the fibre structure, including the pitch and both the holes and the core sizes and shapes. This uniformity is demonstrated in a convincing manner by the very similar photographs of the core region, taken at two ends of the 100m sample of the PBG fibre by the scanning electron microscopy technique (fig17a). This promising result makes it credible to shortly obtain ultra-low loss (<0.2dB/km) transmission of light over a wide range of wavelengths, with non-linearities 100 times lower than in silica.

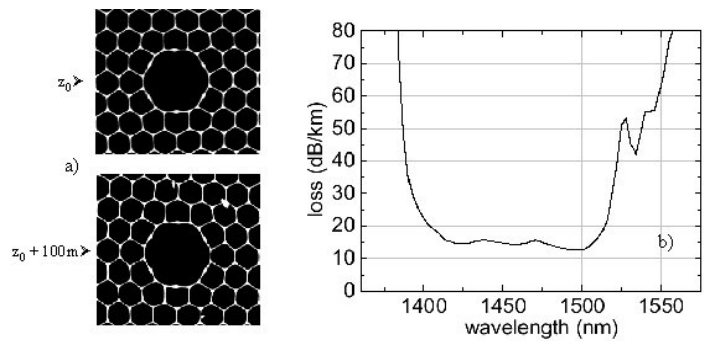

Figure 17: a) SEM photographs of the cross section of a low-loss PBG fibre, taken at the two ends of a 100m sample of fibre ; b) spectral loss of this PBG fibre (*Venkataraman et al. [75]* )
a) photographies par MEB de la section droite d'une fibre BIP à faibles pertes prises aux deux extrémités d'un échantillon de 100m ;
b) atténuation spectrale de cette fibre BIP (Venkataraman et al. [75] )

## V – SOME MOFS-BASED COMPONENTS AND DEVICES

The novel propagation properties of MOFs and PBG fibres are very attractive for number of applications. Furthermore, components derived from those existing in conventional fibres can be realised with MOFs, among which photo-induced Bragg gratings, long period gratings, and tapers. MOFs can also be doped with rare earth for manufacturing amplifiers and lasers. At this point, we must note that for inserting MOFs or MOFs-based components into devices and systems, convenient splices with the upstream and downstream standard fibres must be operated. Obviously, the loss at the connection increases with the mismatch between the distributions of the guided mode into the MOF and into the standard fibre. In order to maintain acceptable connection loss, adequate splicing techniques must be implemented, in particular to prevent the holes from collapsing at high temperature. Heating the fibres into a resistive furnace over a significant length around the connection is a possible mean. However, with a careful local heating provided by the conventional electric arc technique, the mode areas of the two fibres can be adapted as well as possible, so that splice loss as low as 0.3dB can be reached.

In this section, we point out some of the most promising applications of MOFs, and of components based on MOFs and PBG fibres.

### V – 1 rare-earth doped MOFs

Rare-earth doped fibre amplifiers and lasers are based on the interaction of the guided modes at the pump wavelength ($\lambda_p$) and at the signal wavelength ($\lambda_s$) with active rare-earth ions distributed into the core region. The possibility of designing MOFs with mode area significantly different from that of standard step-index fibres combined with a single mode propagation of light over a wide range of wavelengths including both $\lambda_p$ and $\lambda_s$ makes this kind of fibre attractive for designing novel fibre amplifiers and lasers. As above mentioned, the intensity distribution for the guided modes together with its dependence versus the wavelength are considerably different from that of standard step-index fibres. Potentially, the overlap integral of the pump wave and that of the signal over the doped area may be substantially changed. This may result in a modification of the gain curve versus wavelength into MOFs amplifiers. Furthermore, large-mode area lasers are suitable for emitting high power whereas small pump power thresholds and high peak intensities can be obtained in small-area lasers [76, 77].

In spite of these properties, very few experimental works have been published yet. Classically, the rare-earth doped rod that forms the core of the MOF is made by etching away the outer silica cladding from a conventional rare-earth doped fibre preform. When incorporating rare-earth doping ions as well as index dopants (aluminium, germanium, …) into the core, the index of the core is increased by $\Delta n$ and the guidance properties of the MOF can be modified. However, one can consider that the guidance by the holes dominates the guidance induced by the core dopant(s) if the following condition is fulfilled [78]:

$$\Delta n < \frac{n_{silica}^2 - n_e^2(\lambda)}{2 n_{silica}} \qquad (13)$$

The fibre is drawn following the same process as a passive MOF. Then, it is spliced to the output end of a multiplexer used for launching both the pump power and the signal into the doped MOF. It is noticeable that the angular pattern of the spontaneous emission through the cladding of an axially pumped MOF is directly related to the features of the bandgaps of the structure [79]. The study of such a pattern constitutes a rapid mean for characterising this photonic bandgap structure.

The very first $Yb^{3+}$-doped photonic crystal fibre laser is described in [80]. Its efficiency remains somewhat low. Its main interest is that it brings the possibility of soliton fibre lasers at wavelengths around 1μm, thanks to the zero GVD at 730nm and an anomalous dispersion at longer wavelengths. A Nd-doped laser in which the pump radiation illuminates substantial parts of the air hole cladding ("cladding pumped laser") has also been reported [81]. The authors claim that the absorption efficiency is quite good compared to standard double-clad fibres with non-centred cores, thanks to pump mode mixing into the microstructure. However, up to now, the performances of these lasers are not better than that obtained with conventional doped fibres.

More attractive are the potentialities of double clad rare-earth doped fibres based on MOFs [78, 82]. Examples of such fibres are shown on figure 18. The single mode core is surrounded by a large microstructured inner cladding. The outer cladding is made of an arrangement of air holes with a very large air fraction. With such a low index outer cladding, a high NA of the inner cladding is measured: 0.3-0.4 in [78] and above 0.75 in [82]. This is of great interest for increasing the launching efficiency of pump power from multimode laser diodes. Thanks to a small air filling fraction into the inner cladding the propagation into the core can remain single mode with a large effective mode area. A large doped region having a good overlap with the mode can then be used for high power emission. However, as the effective mode area is increased, the NA must be decreased to preserve the single-mode propagation and the mode becomes more and more sensitive to bend loss. The extent of the single mode is then limited in practical applications. Concerning the efficiency of the pump power absorption in the core of double-clad fibres, it has already been demonstrated that it strongly depends on the transverse distribution of power into the modes of the inner cladding and on the core location [83]. One of the advantages of the MOFs fabrication technique is that the core can easily be placed at any point into the inner cladding (fig. 18b). A multimode $Yb^{3+}$-doped laser based on a structure close to that shown on fig. 18b emits up to 3.9W at 1035nm when pumped at 911nm with 20W launched. The maximum slope efficiency is 21% with respect to launched pump power. This efficiency is not higher than that of a large mode area single-clad MOF laser fabricated using the same doped core. Even if the results obtained up to now with rare-earth doped MOFs are not better than those provided by conventional fibres, the novel propagation features and fabrication possibilities of these fibres justify further intensive work.

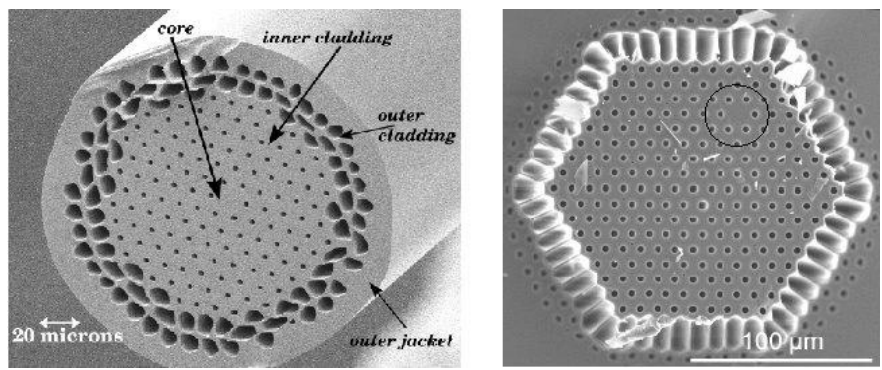

Figure 18: Cross section of two single mode $Yb^{3+}$-doped double-clad MOFs a) MOF from *Furukawa et al. [78]* : inner cladding d=2.7μm Λ=9.7μm diameter=110μm –diameter of the outer cladding=175μm ; b) MOF from *Wadsworth et al. [82]* : inner cladding d=4μm Λ=10.5μm diameter across flat 155μm – overall diameter=500μm
*Sections droites de deux FMAS à double gaine, monomodes, dopées $Yb^{3+}$ a) FMAS proposée par Furukawa et al. [78] : gaine interne d=2.7μm Λ=9.7μm diamètre=110μm –diamètre de la gaine externe=175μm ; b) FMAS proposée par Wadsworth et al. [82] : gaine interne d=4μm Λ=10.5μm diamètre entre méplats 155μm – diamètre hors tout=500μm*

## V – 2 non-linear applications of MOFs

Non-linear optical effects in fibres result from the interaction of optical fields with the glass via the third order non-linear coefficient $\chi^{(3)}$, or Kerr non-linearity. Two features of MOFs are likely to enhance the effects of this non-linearities: the tight confinement of light into high NA small radius core that increases the power density, and the ability of cancelling the GVD at any wavelength between at least 700nm and 1600nm. Many works have already been published on non-linear applications of MOFs. One of the most exciting among them is the continuum generation, as first observed by Ranka et al. [66] and then in many laboratories [67, 84, 85, 86]. The MOF used in the experiments have cores of about 1μm in diameter surrounded by a cladding with an air filling fraction as high as possible. At IRCOM, we have fabricated the MOF shown on the right of figure 19. The core consists in the silica interstice between the three central holes. On the left of figure 19 we present the broad band spectrum generated in a 0.8m-long sample of this fibre when excited by a femto-second Ti:sapphire laser emitting at 838nm [67]. The wavelength generation above 830nm can be mainly attributed to Raman scattering whereas the visible spectrum is due to 4-wave mixing and self-phase modulation. The first demonstration of a broadband continuum generation (450nm to 800nm) in MOFs with non-femto-second pulse pumping

(wavelength=532nm, pulse duration=0.8ns, peak power=400W) is reported in [85]. Applications of MOF supercontinuum generation include broadband sources for biological imaging, for wavelength division multiplexing (WDM) communication systems and for precision frequency metrology [87].

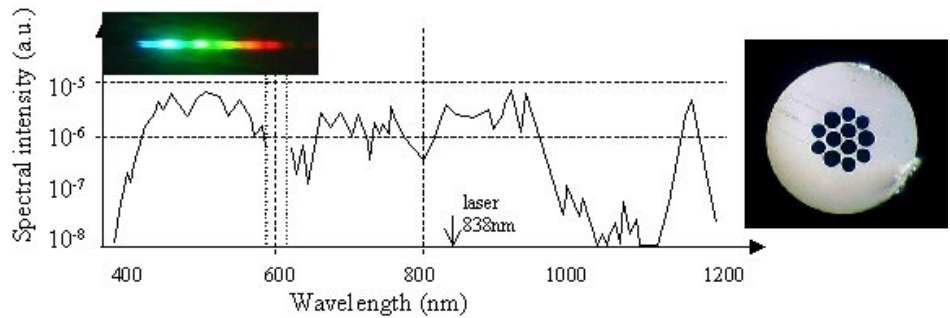

Figure 19: Continuum generation generated by femto-second pulses at 838nm launched into the MOF shown on the right. The visible part of the spectrum (400-600nm) and the infrared part (650nm-1200nm) have been measured with different detectors and cannot be directly compared
*Génération de continuum par des impulsions femtosecondes à 838nm injectées dans la FMAS montrée sur la gauche. La partie visible du spectre (400-600nm) et la partie infrarouge (650nm-1200nm) ont été mesurées avec des détecteurs différents et ne peuvent être comparées directement*

Different other experiments dealing with highly non-linear MOFs have also been reported in the literature. Cross-phase modulation can be used to achieve all-optical switching, as it is demonstrated near 1550nm in [88]. This phenomena is also exploited by Lee et al. for demonstrating a tunable WDM wavelength converter into a MOF [89]. Wavelength conversion over a 20nm tuning range was achieved with negligible pulse width variation. A combination of soliton propagation and of stimulated Raman scattering into a 15cm-long tapered MOF results in a self-soliton frequency shift, as shown in [90]. An optical parametric oscillator has been realised using 4-wave mixing into a 2.1m long MOF [91]. In this fibre, the GVD is zero near 745nm. When pumping at 752nm, a resonant signal tunable over 40nm is obtained around 765nm at the output of the oscillator. Brillouin scattering into MOFs is investigated in [92]. For example, a Brillouin laser based on a 73m long MOF has been realised. The threshold is observed when the fibre is pumped by 125mW at 1552nm.

In order to obtain substantial non linear effects in low-density media such as gases, a long interaction length with the laser light and a high intensity are required. PBG fibres can be used to fullfil these conditions due to the fact that the core is a hole that can be filled with the gas and can efficiently guide light over a long length. This concept has been implemented by Benabid et al. to achieve efficient stimulated Raman scattering in hydrogen gas [93]. Hydrogen gas under pressure was injected into the 15μm-diameter core of a 30cm-long sample of PBG fibre (see section IV). The threshold for Stokes generation was observed for pulses of the laser source (duration=6ns, λ=532nm) having an energy as low as 800±200nJ. The anti-Stokes generation threshold appeared for energies equal to 3.4±0.7μJ. With pulse energy of only 4.5μJ, the pump to Stokes conversion efficiency reached 30±3%. This energy is two orders of magnitude lower than that previously reported. This considerable progress probably opens the way to a new area in gas-based non-linear optics.

## V – 3 MOF-based components using Bragg gratings, long period gratings, and tapers

By combining the influence of the holes and the action of Bragg gratings or long period gratings (LPG) on the propagation of light into MOFs, numerous novel components achieving useful functionalities, tunable or not, can be designed.

To allow the inscription of Bragg gratings into the core of a MOF, the rod placed in the preform for constituting the core must be photo-sensitive, i.e. germanium-doped. This doping induces an increase of the index. To preserve the predominant guidance by the holes, the condition expressed in (13) must be verified. The very first demonstration of optical fibre grating written in a MOF has been reported in [94]. A 1µm Ge-doped core was set in the centre on the MOF (d=2µm, Λ=10µm). The fibre was loaded with deuterium and then exposed over about 4cm to 242nm-light through a conventional phase mask. In spite of the very low overlap of the fundamental mode with the grating (~0.05), a Bragg reflection of about 50% at $\lambda_B = 2n_e.\Lambda_B = n_e.\Lambda_{Mask}$ was observed ($\Lambda_B$ is the Bragg wavelength and $\Lambda_{Mask}$ is the period of the mask). LPGs can also be written by this technique [94]. However, it is possible to create LPGs in Ge-free MOFs. Kakarantzas et al. used a $CO_2$ laser technique but they periodically completely collapsed the holes by each laser heat treatment [95]. This can result in an increase of the loss due to a poor guidance of the mode and some applications that necessitates to use a flow of fluids in the holes (see below) are no longer allowed. Humbert and his colleagues, in collaboration with us, have shown that electric discharges provided by a commercial splicer, periodically applied on the MOF along the z-axis, allow the realisation of an efficient LPG with only very slight deformations of the fibre [96]. The insertion loss is lower than 0.5dB and the isolation is better than 15dB at the operating wavelength. Furthermore, the sensitivity of $\Lambda_B$ to the temperature remains as low as 9pm/°C. By means of heating techniques, tapered MOFs can also be manufactured [97].

The holes of MOFs can be filled with active material provided that they are sufficiently large. Thus, light propagating into such a MOF can be manipulated if there is an efficient interaction between the field and the material. The first way to achieve this requirement is to spread the mode into the cladding by tapering the fibre. The second way is to provoke coupling between the core and the cladding modes by means of a grating. Starting from this concept, Eggleton et al. propose a large variety of tunable devices [98]. The MOF consists in a Ge-doped core (diameter=8µm, $\Delta n \sim 5.10^{-3}$) surrounded by 6 large air holes (30µm in diameter) in a 125µm outer diameter silica jacket (fig.20a). The active material, that can be a microfluid or a monomer mixture, is drawn up into the holes by vacuum applied at one end of a short sample of the MOF. The monomer mixture forms a polymer with the desired refractive index by UV exposure of the fibre. The polymer may be chosen to have a refractive index highly dependent on the temperature. When this refractive index becomes higher than that of silica, the mode field refracts into the polymer and this results in a dramatic increase of the loss into the core. To enhance the interaction between the field and the polymer this experiment in realised in a tapered MOF (fig.20b). Based on this principle, an electrically tunable attenuator device with insertion loss lower than 0.8dB (including the splice loss with standard fibres at each end) has been demonstrated (fig.20c) [99]. The dynamic from 40°C to 120°C is higher than 30dB as shown on the curve of figure 20d.

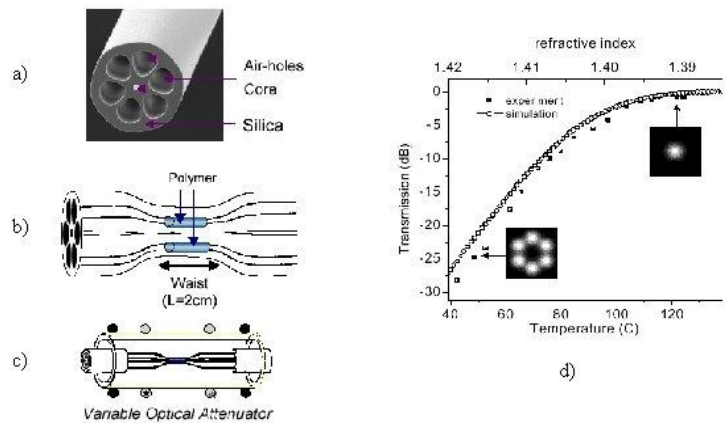

Figure 20: a) MOF with 6 large air holes ; b) tapered MOF with polymer ; c) tunable attenuator device ; d) attenuation of the device shown in c), versus the temperature (in the insets are shown the distribution of the field into the fibre at low (resp.high) temperature, i.e. when the index of the polymer is higher (resp. lower) than that of silica) (*Eggleton et al. [98]*)
*a) FMAS avec 6 gros trous d'air ; b) FMAS amincie contenant un polymère ; c) atténuateur variable ; d) atténuation du dispositif montré en c), en fonction de la temperature (dans les encarts on montre les distributions du champ dans la fibre à basse (resp. haute) température, c.à.d. quand l'indice du polymère est plus élevé (resp. moins élevé) que celui de la silice) (Eggleton et al. [98])*

The transmission spectrum of a LPG written in the core of such a MOF can be tuned thanks to the same principle. The cladding resonances can be wavelength shifted over 150nm around 1500nm when heating the polymer from 35°C to 120°C [98]. This allows the easy realisation of a tunable grating filter. When only certain opposite holes are selectively filled with polymer, a high birefringence can be created into the fibre [100]. This birefringence is tunable by monitoring the temperature. By filling two opposite holes with the polymer, a birefringence from $3.10^{-4}$ to $6.10^{-4}$ has been measured when the temperature is varied from 150°C to 20°C. Furthermore, the leakage of the mode into the polymer results in a differential loss versus the polarisation. It increases from 0.4dB to 1.3dB when the temperature varies from 120°C to 30°C. A variable optical polarizer can be realised taking advantages of this feature.

At last, when the infused material is a microfluid, it can be moved back and forth along the MOF by applying a pressure or a depression at one end of the MOF. This can also be achieved by heating or cooling the air boxed up in the holes. This gives a mean of temporarily inhibiting the action of the material if necessary in some applications.

## V – 4 Other developments related to microstructured fibres

As shown above, in addition to the direct application of the specific propagation properties of MOFs, the three main research axis on these fibres deal with rare-earth doped MOFs, non-linear effects in MOFs and novel components based on the interaction of the guided field with a material infused in the air-holes. Besides, some other fields are worth noting. Generally speaking, they concern new structures of MOFs, MOFs made of material different from silica, or specific applications of MOFs or PBG fibres.

Among the new possible structures, some work has been devoted to multicore MOFs. Indeed, thanks to the technology of fabrication, it is easy to design a MOF with two or more cores. The coupling of light from one core to a neighbouring one along a dual core MOF has been experimentally studied, as shown on figure 21 [101]. By means of a 3D full-vectorial BPM, it has also been theoretically shown that no significant polarisation coupling occurs in such a fibre, meaning that the structure operates in polarisation preserving mode [36]. As it is difficult to separate light from each core in practical applications, it seems hazardous to think of easily using such fibres as couplers or multiplexers. At this date, the main impact of these studies is to give information on the degree of modal confinement into the cores.

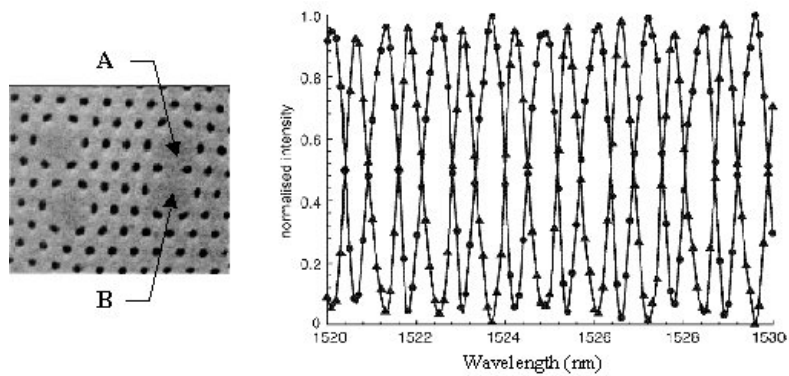

Figure 21: Intensity transmitted by each core A and B of the MOF shown on the left when light is launched in the core A, as a function of wavelength (d=0.95μm, Λ=2.8μm, intercore spacing = √3Λ)  (*Mangan et al. [101]* )
*Intensité transmise par chaque cœur A et B de la FMAS montrée sur la gauche quand la lumière est injectée dans le cœur A, en fonction de la longueur d'onde (d=0.95μm, Λ=2.8μm, espacement entre les coeurs =  √3Λ)  (Mangan et al. [101] )*

Taking into consideration the demonstrated potentialities of air-silica MOFs, different laboratories have started investigations on MOFs made of materials different from silica. Some polymer MOFs have already been fabricated [17,102]. Extrusion, polymerisation in a mould or injection moulding are techniques for making the proform that allow to design cross-sections much more difficult to realise with the stacking technique of silica capillaries. Doping the structure with high concentrations of additive for enhancing non-linearities or magneto-optics effects is also easier. However, because of the much higher loss in polymer than in silica, polymer MOFs are preferentially devoted to short length components. Other studies deal with compound glass MOFs, such as chalcogenide fibres. The feasibility of a MOF made of gallium lanthanum sulphide (GLS) glass is demonstrated in [103]. This material is transparent at wavelengths up to 5µm, has a refractive index as high as 2.3 and can be doped with high concentration of rare-earth (over 10% in weight). The fabrication technique is analogous to that used for silica MOFs. Because the fibres are made from a single material, the serious problems encountered when manufacturing conventional compound glass fibres made of materials with different physical properties (one constituting the core and one constituting the cladding) are eliminated. High quality chalcogenide MOFs are not available yet but a notable research effort is devoted to improve this kind of fibres. It is justified by their potential applications in low-loss transmission in the 1 to 5µm wavelength range and in non-linear optics.

At last, let us note a specific use of PBG fibres, dealing with the levitation and the guidance of particles in the hollow core by means of the force of radiation pressure. It allows the non-intrusive manipulation of microscopic objects or particles, very attractive in many areas such as biology, chemistry or atomic physics. The transportation of micro-sized objects by a focused laser beam in free space has already been demonstrated but only over a distance of about 1mm. The natural use of a hollow-core fibre capillary can enhance this distance to only few millimetres because of the strong unavoidable loss of this guide. A PBG silica fibre with a 20µm hollow core diameter and an air filling factor of 75% has been used to guide polystyrene particles over 150mm at a speed of 1cm/s with a laser power as low as 80mW [104]. At the laser wavelength (514nm) the loss is ~5dB/m. With a loss as low as 13dB/km (see fig.17 and [75]) the possible guidance length could be increased to almost 130m. This opens new prospects for all applications requiring the transportation of micro-objects.

# VI - CONCLUSION

Microstructured optical fibres (MOFs) have been intensively studied in many laboratories all over the world for the last few years. Hundreds of papers and contributions in conferences dealing with theoretical and experimental studies of MOFs have been published in the two first years of the 21th century. There are two main motivations for this tremendous interest. First of all, the novel propagation properties of these fibres, unreachable with conventional fibres, open the way for a large number of applications in various domains such as optical telecommunications components and systems, optical sources, frequency metrology, high resolution spectroscopy, sensors, etc. A second reason is that their fabrication process allows the design of new geometries of structures that cannot be obtained with the classical MCVD or OVD techniques.

In this paper, we have reviewed the recent progress in both theoretical and experimental studies of MOFs and PBG fibres. Not less than seven methods for modelling these fibres are available. It is obvious that, due to the complex structure of actual MOFs, the mode-finders methods necessitate a fine discretisation of the cross-section and are consequently time and memory consuming. In some cases, approximate methods such as the effective V-model or the equivalent average index method (EAIM) may provide rough but sufficient values of the effective index of the propagating modes. It is also sometimes possible to take advantages of symmetries of the structure. In the case of fibres or components which index profile varies along the z-axis, the beam propagation method (BPM) may be implemented although it necessitates a very long computation time for modelling a significant length of fibre. In this paper, the advantages and the drawbacks of the main methods have been pointed out. In spite of a lack of data from the literature, we have discussed the domains of validity of the main ones. In particular, we have shown that the characteristics computed in our laboratory with the finite element method, the EAIM and the BPM and are in good agreement with the measured ones.

Very significant improvements have been recently obtained concerning the propagation properties of MOFs. The endlessly single-mode propagation into properly designed MOFs, particularly attractive for many applications, has been widely verified. The propagation loss has been dramatically reduced to values as low as 0.61dB/km at 1550nm. This now allows to consider the practical use of these fibres in long haul transmissions. The chromatic dispersion can be adjusted in order to fit with the needs of particular applications: flattened dispersion for high bit rate telecommunications or dispersion cancelled at quite any wavelength for non-linear optics. However, this feature remains very sensitive to the opto-geometrical parameters of the fibre. Thanks to the fabrication process, highly birefringent MOFs can be easily manufactured. In particular, in our laboratory, we have designed and measured a single-mode highly birefringent MOF with a beat length as short as 0.4mm. Another important feature, the effective-mode area, can be managed in MOFs, from just few µm$^2$ for enhancing non-linear effects to tens of µm$^2$ for propagating optical power with reduced non-linear Kerr effects.

All the novel propagation properties listed above have given rise to number of developments. Amplifiers and lasers based on double-clad rare-earth doped MOFs are promising components for efficient amplification using low-cost high-power multimode pump laser diodes. This is due to the very large numerical aperture (up to 0.75) that can be reached for the inner cladding. Another advantage is that the outer polymer cladding is no more necessary. At this date, the obtained performances are not as high as anticipated but further improvements in this domain are expected. On the contrary, in non-linear optics, a lot of exciting results have already been obtained. They arise from both the high power density of light in small core MOFs and the cancellation of the group velocity dispersion at the operating wavelength. One of the most significant is the continuum generation when launching short pulses in a suitable sample of a MOF. For example, we have manufactured a MOF in which a femtosecond pulse around 838nm is spectrally enlarged over more than 700nm. Numerous other devices based on non-linear optics into MOFs are reviewed in this paper. At last, MOFs with microfluids or polymers infused into the holes have been demonstrated to constitute the key element for achieving new functionalities. Bragg gratings, long period gratings and tapers are used to increase the interaction between the guided wave and the material into the holes. Novel tunable components based on these structures, such as attenuators or polarisers, have been successfully realised.

Few years ago, we have experimentally demonstrated the ability of guiding light into the low index core of a Bragg fibre manufactured with the MCVD technique. More recently amazing progress have been reported on MOFs operating on a true photonic band gap effect and able to guide light in hollow core. The loss of these PBG fibres has been dramatically reduced to 13dB/km around 1500nm. This promising result makes it credible to shortly achieve ultra-low loss (<0.2dB/km) transmission of light over a wide range of wavelengths, with non-linearities 100 times lower than in silica. Raman scattering into a hydrogen-filled core with very low pump level and particle guidance into the hollow core over a long distance using the radiation pressure constitute other exciting applications of PBG fibres.

In conclusion, it is obvious that the microstructured optical fibre technology opens the way to many advanced applications that were not reachable with conventional fibres. Although this technology is still in its infancy, MOFs have already proven number of their potentialities. Considering the massive effort made in numerous major research groups for improving MOFs and PBG fibres and for applying them in various fields of physics, we can expect considerable new progress in the very near future. We are at the beginning of a new era in the domain of guided optics.

Acknowledgements: A large part of the work achieved at IRCOM on MOFs was supported by funding from Alcatel. The authors are also grateful to the Centre National de la Recherche Scientifique and to the Region Limousin for their financial support through research programs. Ambre Peyrilloux is currently supported by an ANRT-CIFRE convention.
The authors would like to thank the Linear Microwave Devices Group of IRCOM for placing at their disposal its FEM software, and Gilles Renversez from Institut Fresnel (UMR CNRS 6133) for helpful discussions on the multipole method. They acknowledge Pierre Sansonetti, Laurent Gasca, Gilles Melin, Lionel Provost, Isabelle Bongrand and Cyril Rullier from Alcatel Research and Innovation in Marcoussis for numerous invaluable exchanges and for providing some of the tested fibres. Thank you to Thierry Chartier and Abdelrafik Malki from CORIA (UMR CNRS 6614) for fruitful collaborations.
The last but not the least, are the acknowledgements intended for colleagues from IRCOM involved in some parts of this work: Jean-Marc Blondy and Jean-Louis Auguste for their participation in drawing fibres, Frédéric Louradour and Alain Barthélémy for actively contributing in characterising the fibres.